\documentclass[aps,prl,twocolumn,showpacs,floatfix]{revtex4-1}

\usepackage{graphicx,epsfig}

\begin{document}

\title{A homoclinic tangle on the edge of shear turbulence}

\author{Lennaert van Veen}
\email{lennaert.vanveen@uoit.ca}
\thanks{Supported by NSERC Grant nr. 355849-2008}
\affiliation{Faculty of Science, University of Ontario Institute of Technology, 2000 Simcoe St. N., Oshawa, L1H 7K4
Ontario, Canada}

\author{Genta Kawahara}
\email{kawahara@me.es.osaka-u.ac.jp}
\thanks{Supported by JSPS Grant-in-Aid for Scientific Research}
\affiliation{
Graduate School of Engineering Science,
Osaka University, 1-3 Machikaneyama, Toyonaka, Osaka 560-8531, Japan}

\date{\today}

\begin{abstract}
Experiments and simulations lend mounting evidence
for the edge state hypothesis on subcritical transition 
to turbulence, which asserts that simple states of 
fluid motion mediate between laminar and turbulent 
shear flow as their stable manifolds separate the 
two in state space. In this Letter we describe a flow 
homoclinic to a time-periodic edge state. Its 
existence explains turbulent bursting through the 
classical Smale-Birkhoff theorem. During a burst, 
vortical structures and the associated energy 
dissipation are highly localized near the wall, in 
contrast to the familiar regeneration cycle.
\end{abstract}


\maketitle

\begin{figure}[t!]
\begin{picture}(300,410)
\put(-10,-42){\epsfig{file=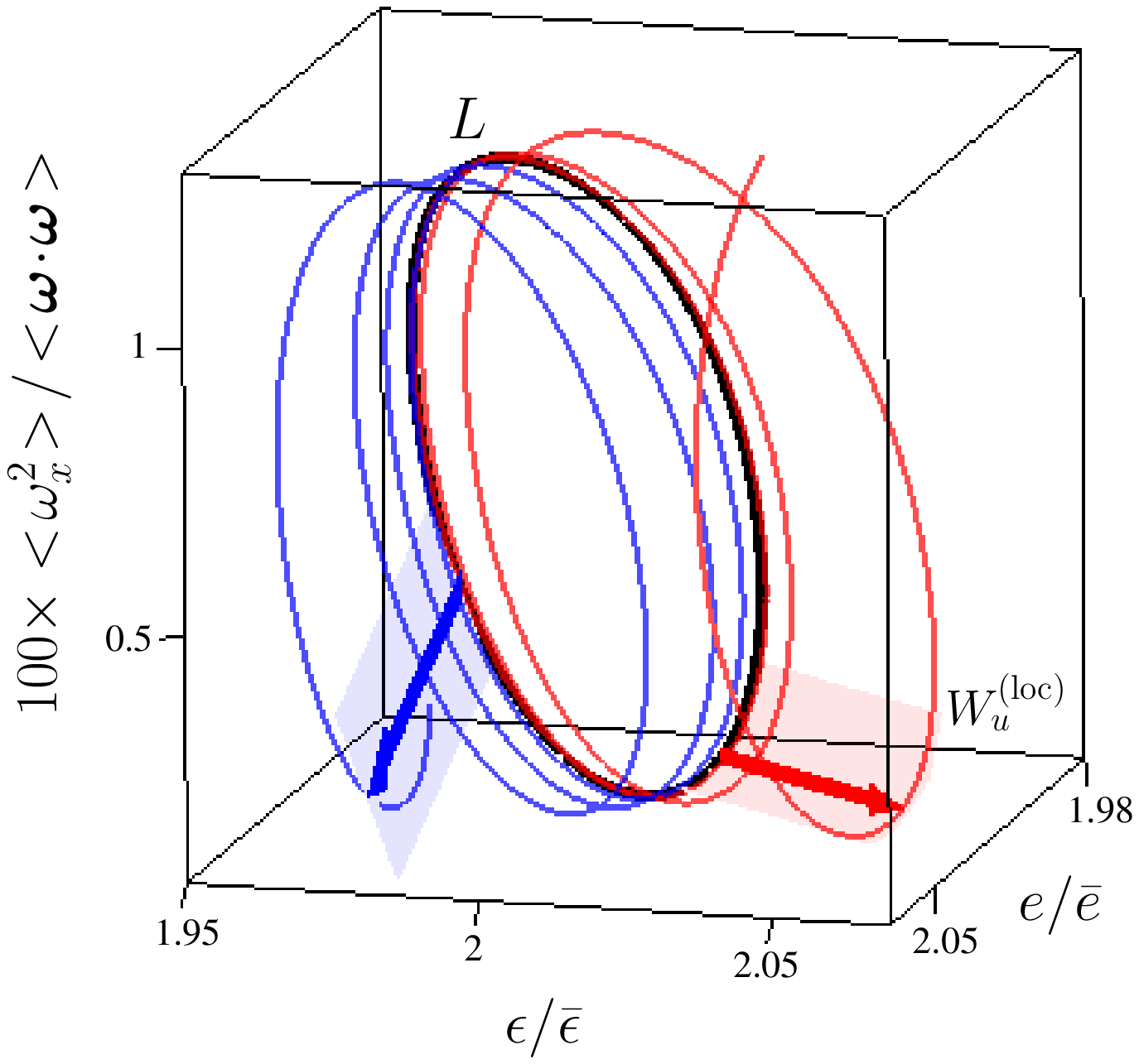,width=0.6\textwidth}}
\put(10,190){\epsfig{file=Fig_1a.eps,width=0.42\textwidth}}
\end{picture}
\caption{Top: projection of the homoclinic orbit onto energy input and dissipation rate, normalized by their
value in laminar flow,
$\overline{e}$ and $\overline{\epsilon}$.
The piece of orbit leaving the edge state, $L$, is shown in red and the one approaching it in
blue. In the background, the PDF of transient turbulence is shown in gray scale.
The labels a--f correspond to the snap shots in Fig.\ref{physical_space}.
The inset shows
the input (red) and dissipation (green) rate of energy
as a function of time, normalized with $T$, along part of the homoclinic orbit.
Bottom: close-up of the UPO, $L$, with the homoclinic. The red and the blue arrow denote the projected local 
unstable manifold and the
projected leading stable subspace, respectively. On the axes are the normalized 
energy input and dissipation rate and the
mean square streamwise vorticity normalized by the mean square total vorticity, where the mean is 
taken over the flow unit.
\label{homoclinic_orbit}}
\label{phase_space}
\end{figure}

\noindent{\bf Introduction}\hspace{10pt} In recent years, the open problem of subcritical transition to turbulence in 
shear flows has seen a surge of interest, sparked by a rapidly increasing ability to perform numerical simulations as 
well as by a string of novel applications of dynamical systems theory. The dynamical systems approach
advocates the idea, that transitional shear flow is regulated by special solutions with a relatively simple spatial
structure. They may be travelling waves, time-periodic solutions or even solutions chaotic in time.
Such states live in between a stable laminar state and stable, or meta-stable, turbulence.
In simulations, these special states can be identified by a shooting algorithm in which initial data are
iteratively refined to yield a flow which neither laminarizes nor becomes turbulent, but instead lingers
on the ``laminar-turbulent boundary'' \cite{itan,*itan2}. States on this boundary are necessarily unstable
and consequently they
can only be observed as transient effects in experiments \cite{[{With pipe flow as celebrated example: }][{}]hof1}. 
Their relevance to the
transition process lies mainly in their stable and unstable manifolds in state space. These determine how
the fluid behaves as it transitions from near-laminar to turbulent states and vice versa.\\
A particularly interesting situation arises when the laminar-turbulent boundary is formed - at least locally - by the
stable manifold of a traveling wave or periodic solution, which is then called an ``edge state''.
Edge states have now been computed for flow in channels as well as
pipes, with various numerical schemes and spatial discretizations, and there is growing consensus that
they are a robust feature of subcritical shear flow \cite{[{See, e.g., }][]eckh}.\\
Logically, the next step in this analysis would be to study the stable manifolds of the edge states.
Knowledge of the geometry of these manifolds would open the door to the application of control
techniques, aiming at a forced 
laminarization
of the flow. Indeed, some results to this effect have
been obtained using linearization about an edge state \cite{kawa05}. Little, if anything, is know
about the global, nonlinear structure of these separating manifolds, greatly reducing the usefulness
and predictive power of the edge state hypothesis. The direct study of a separating manifold is
hard if not impossible, owing to its high (formally {\em infinite}) dimensionality.
In the current Letter, we commence by studying
the two-dimensional {\em unstable} manifold of an edge state in plane Couette flow, using a novel computational
algorithm. We find that it contains an orbit which returns to the edge state along its stable manifold.
Through the classical Smale-Birkhof theorem \cite{smale}, the presence of this
homoclinic
orbit implies 
the existence of an intricate tangle of the stable and unstable manifolds and chaotic dynamics which 
manifests itself as irregular turbulent bursting.\\
\noindent{\bf Transitional plane Couette flow}\hspace{10pt}
We consider plane Couette flow at a Reynolds number of $Re=400$ in
the minimal flow unit of dimensions $L\times W\times H$, where
$Re$ is based on
half
the velocity difference between the two walls, $U/2$,
and
half
the wall separation, $H/2$.
The streamwise and spanwise periods are
$(L/H,W/H)=(2.76,1.88)$~\cite{hamilton}. We used resolutions of $16\times 33 \times 16$
as well as $32\times 33\times 32$ grid points
in the streamwise ($x$),
wall-normal ($y$)
and spanwise ($z$) directions, respectively,  and checked that the behaviour is
qualitatively the same.\\
In this computational domain, no Nagata \cite{nagata}
steady solution exists
\cite{jimenez}.
Instead,
the edge state in this flow is a time-periodic variation of the laminar
flow, which shows weak, meandering streamwise streaks \cite{kawa,kawa05}.
This gentle Unstable Periodic Orbit (UPO) has a single unstable Floquet
multiplier and thus its unstable manifold has dimension two and its stable manifold 
has codimension one in state space. Both the UPO and its unstable manifold
are contained in a subspace invariant under the spatial symmetries given by reflection in the mid plane,
followed by a streamwise shift over $L/2$, and reflection in the streamwise
and spanwise direction, follow by a spanwise shift over $W/2$.
Consequently, we can impose these symmetries on the solutions to reduce the number
of degrees of freedom, without placing artificial restrictions on the 
fluid motion. At the higher resolution, the number of degrees of freedom in the
simulations is about $11,000$.\\
\noindent{\bf Homoclinic orbit computation}\hspace{10pt} The unstable manifold
of the UPO can be computed using multiple-shooting orbit continuation \cite{veen}. 
Essentially,
this algorithm produces a sequence of orbits, contained in the manifold,
by arclength continuation of a boundary value problem in time.
Because of the extremely sensitive dependence on initial conditions in turbulent
Navier-Stokes flow, we need to compose the orbits of multiple segments, each not
much longer than the decorrelation time, i.e. the time scale of exponential divergence 
of
turbulent
states of fluid motion which are initially close.
The results presented below use up to six shooting intervals, the integration time
on each interval staying below two times the period $T$ of the UPO and below
five times the decorrelation time.\\
As we compute a sequence of orbits, it may happen that it converges
to a homoclinic orbit, which separates the unstable manifold into two components.
Such an orbit is shown in
Fig. \ref{phase_space}(top). In this projection on energy input and
dissipation rate, the UPO is the tiny loop labelled $L$. 
In the background we have plotted the Probability Density 
Function (PDF) of transient turbulence. In the transition to turbulence the 
homoclinic orbit overshoots the maximum of the PDF, then passes close to it
on the way back to the UPO. Clearly, the shape of the PDF
agrees well with that of the homoclinic.\\
We generically expect
the homoclinic orbit to approach the UPO along the least stable subspace
of the latter, also called the leading stable subspace. This is visibly the case in Fig \ref{phase_space}(bottom), which shows a
close-up
of the UPO, along with the local unstable manifold in red and the
local leading stable subspace in blue, now with a third axis
showing
the mean square streamwise vorticity.
How well the depicted orbit approximates a homoclinic connection can be 
quantified by measuring two distances: 
one from the end point of the computed orbit
to the leading stable subspace at a point on the UPO and one to this point along
the leading stable subspace.
We found these distances to be of
order $10^{-7}$ and $10^{-6}$ in the energy measure, respectively.
To test the robustness of this result, we have recomputed the homoclinic
with a smaller integration time step and a varying number of shooting intervals.
\begin{figure*}
\begin{center}
     \begin{minipage}{.3\linewidth}
\begin{picture}(0,0)
\put(0,3){a}\put(2,5){\circle{10}}
\end{picture}
      \includegraphics[clip,width=\linewidth]
      {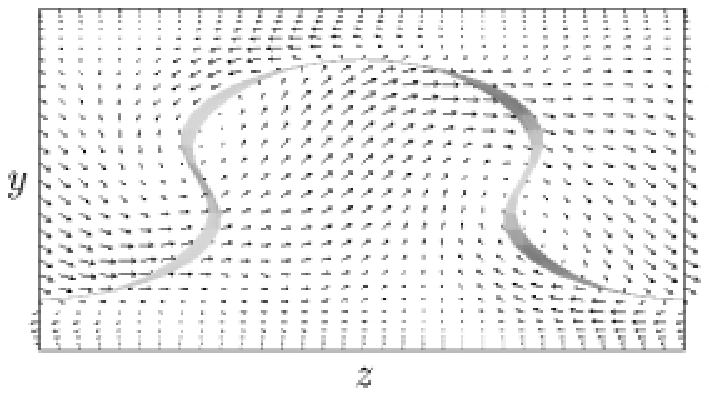}
     \end{minipage}
     \begin{minipage}{.3\linewidth}
\begin{picture}(0,0)
\put(0,3){b}\put(2,5){\circle{10}}
\end{picture}
      \includegraphics[clip,width=\linewidth]
      {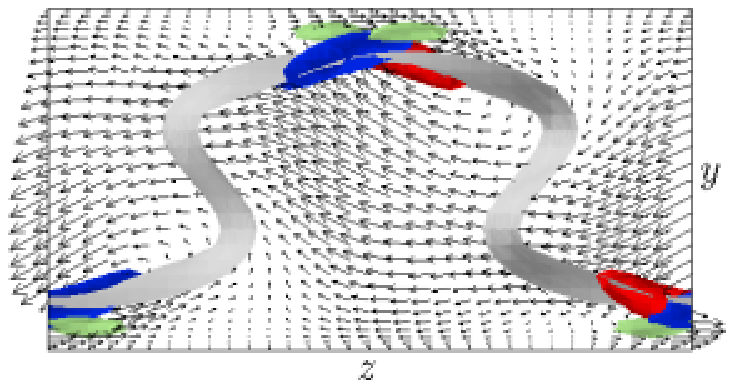}
     \end{minipage}
     \begin{minipage}{.3\linewidth}
\begin{picture}(0,0)
\put(0,3){c}\put(2,5){\circle{10}}
\end{picture}
      \includegraphics[clip,width=\linewidth]
      {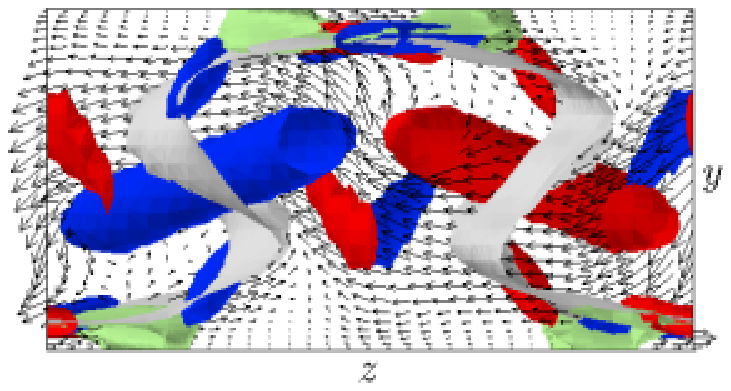}
     \end{minipage}\\
\hspace*{2ex}
     \begin{minipage}{.3\linewidth}
      \includegraphics[clip,width=\linewidth]
      {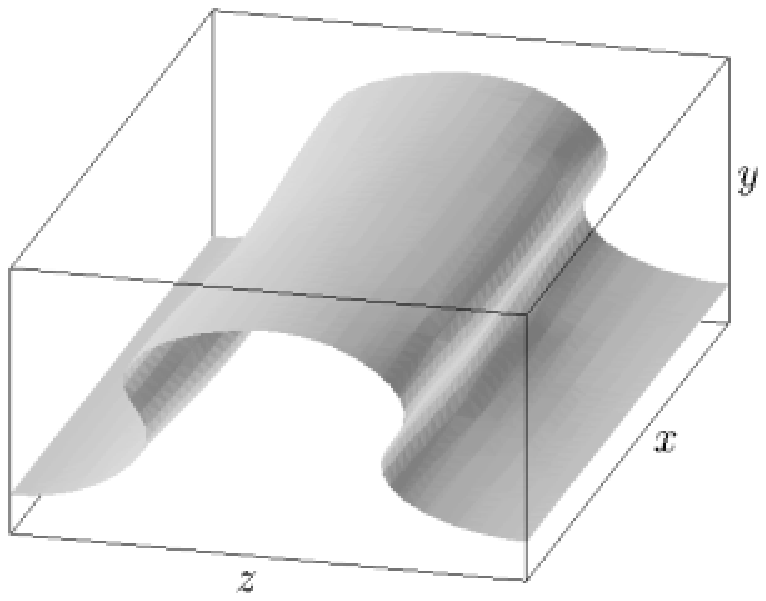}
\begin{picture}(0,0)
\end{picture}
     \end{minipage}
     \begin{minipage}{.3\linewidth}
      \includegraphics[clip,width=\linewidth]
      {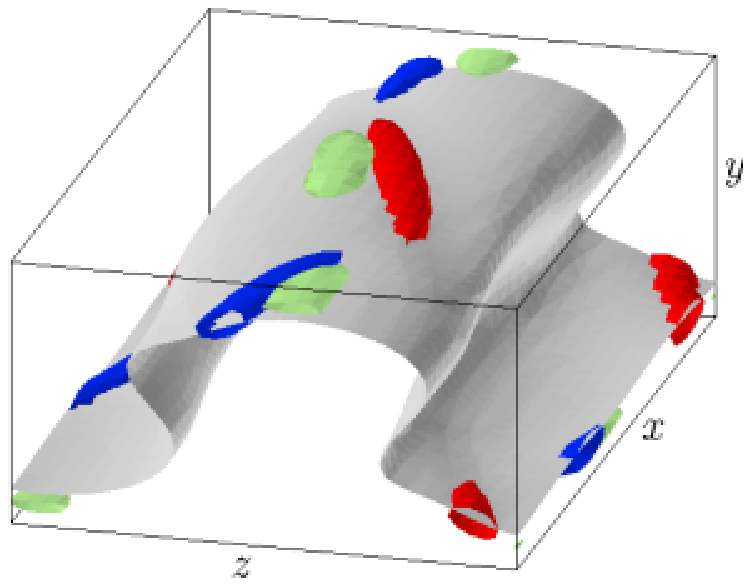}
\begin{picture}(0,0)
\end{picture}
     \end{minipage}
     \begin{minipage}{.3\linewidth}
      \includegraphics[clip,width=\linewidth]
      {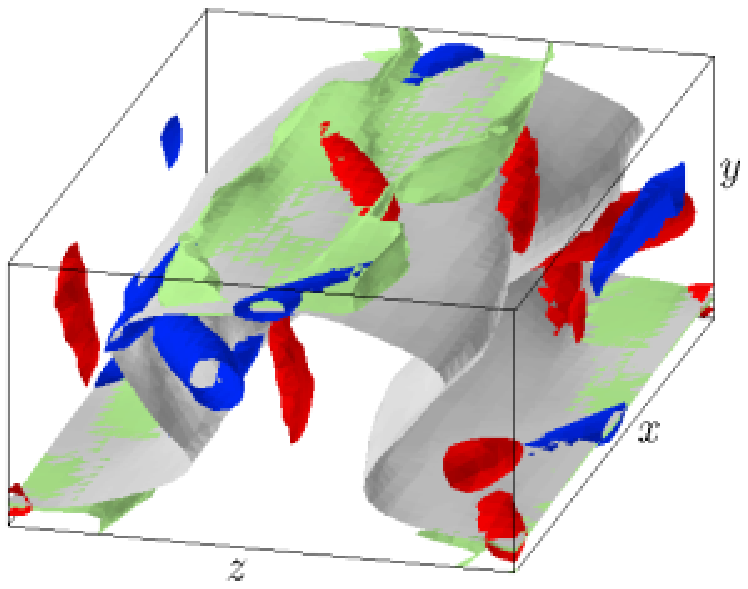}
\begin{picture}(0,0)
\end{picture}
     \end{minipage}\\[5mm]
     \begin{minipage}{.3\linewidth}
\begin{picture}(0,0)
\put(0,3){d}\put(2,5){\circle{10}}
\end{picture}
      \includegraphics[clip,width=\linewidth]
      {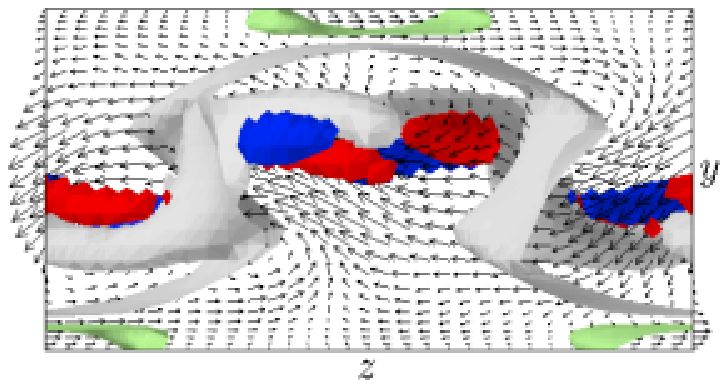}
     \end{minipage}
     \begin{minipage}{.3\linewidth}
\begin{picture}(0,0)
\put(0,3){e}\put(2,5){\circle{10}}
\end{picture}
      \includegraphics[clip,width=\linewidth]
      {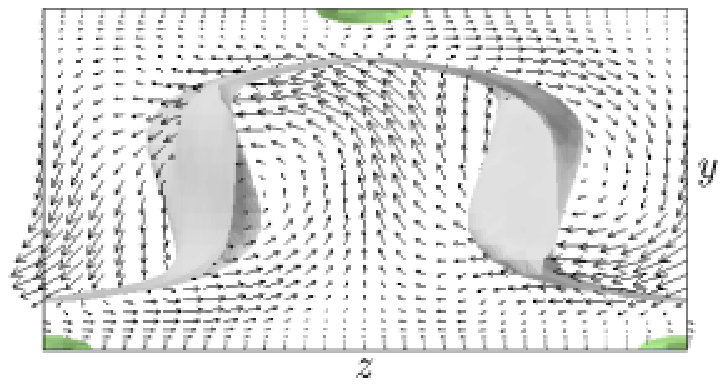}
     \end{minipage}
     \begin{minipage}{.3\linewidth}
\begin{picture}(0,0)
\put(0,3){f}\put(2,5){\circle{10}}
\end{picture}
      \includegraphics[clip,width=\linewidth]
      {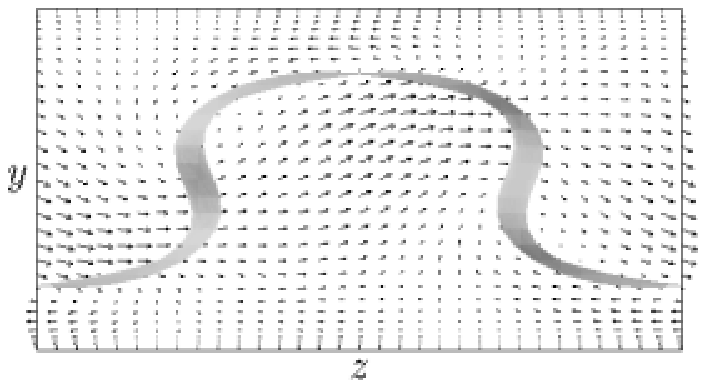}
     \end{minipage}\\
\hspace*{2ex}
     \begin{minipage}{.3\linewidth}
      \includegraphics[clip,width=\linewidth]
      {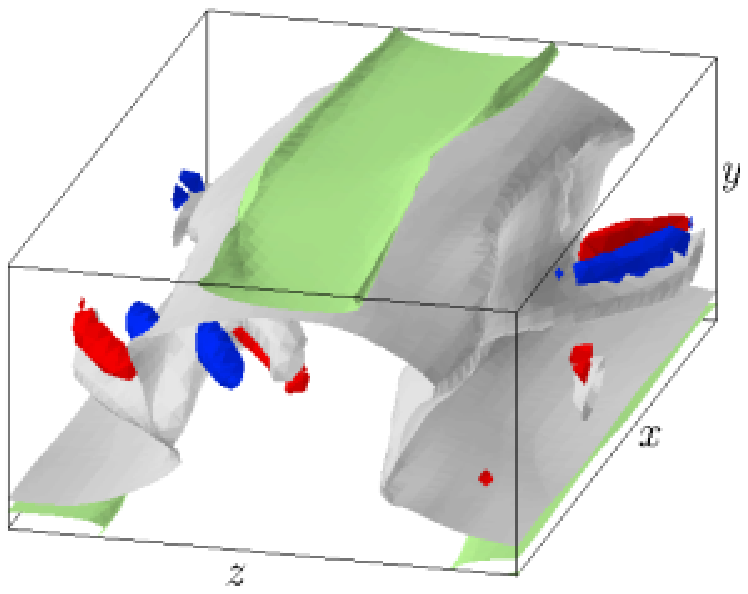}
\begin{picture}(0,0)
\end{picture}
     \end{minipage}
     \begin{minipage}{.3\linewidth}
      \includegraphics[clip,width=\linewidth]
      {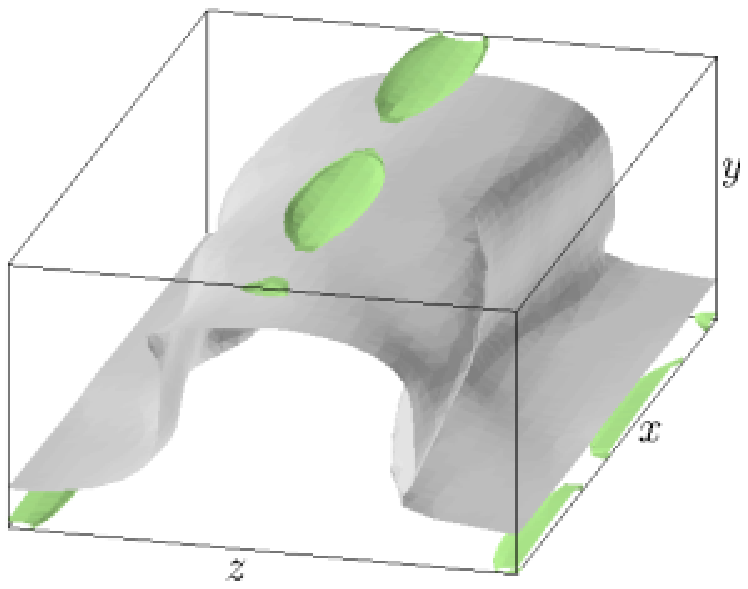}
\begin{picture}(0,0)
\end{picture}
     \end{minipage}
     \begin{minipage}{.3\linewidth}
      \includegraphics[clip,width=\linewidth]
      {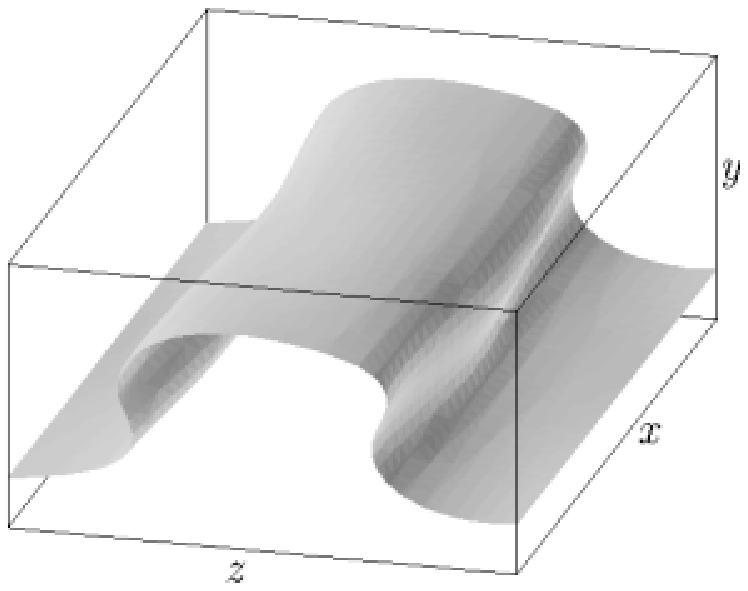}
\begin{picture}(0,0)
\end{picture}
     \end{minipage}\\
\caption{
Visualizations of flow structures in one periodic box $L\times W\times H$
in six phases on the homoclinic orbit, labeled as in Fig.~\ref{phase_space}.
In each phase the structures are viewed in two distinct directions.
Gray corrugated isosurfaces of the null streamwise ($x$) velocity
represent streamwise streaks.
Red and blue
objects are isosurfaces at
$0.2(U/H)^2$
for the second invariant of a velocity gradient tensor,
and denote the vortex tubes of the positive (clockwise)
and negative (counter-clockwise) streamwise-vorticity component.
Green isosurfaces show the local energy dissipation rate at 
$20$ times the value $\bar{\epsilon}$ in laminar flow.
Cross-stream ($y,z$) velocity is shown in the mid plane $x=L/2$.
By the flow symmetry it is related to that at $x=0$ by a reflection
in the spanwise direction.
\label{physical_space}}
\end{center}
\end{figure*}
\begin{figure}[t!]
\begin{center}
      \includegraphics[clip,width=.8\linewidth]
      {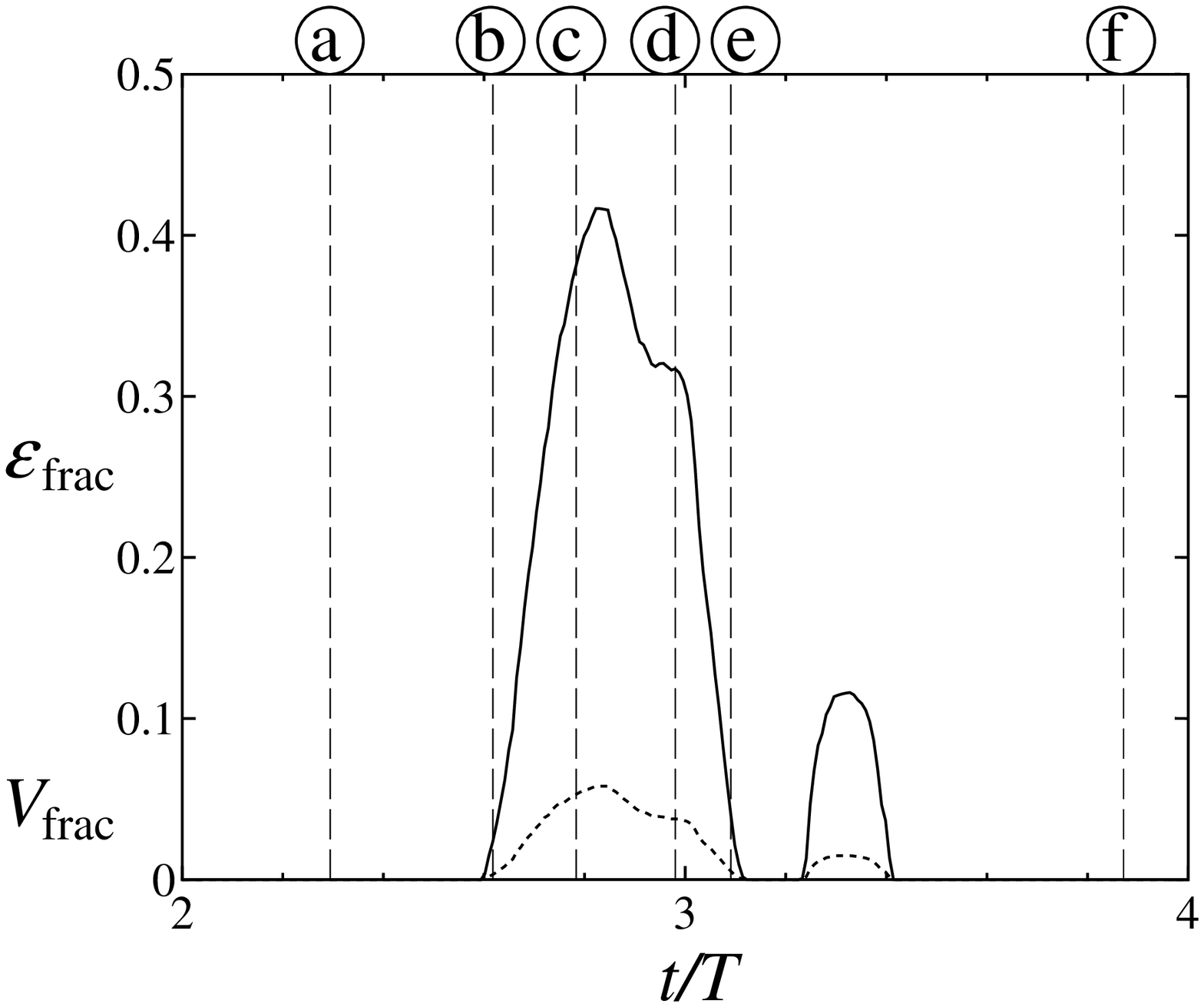}
\caption{
Temporal variation of energy-dissipation (solid curve) and volume (dotted curve) fractions
in the region
bounded by the green surfaces in Fig.~\ref{physical_space},
where the local dissipation rate is greater than 
$20$ times the value in laminar flow.
Time $t$ is normalized with 
the period of the gentle UPO, and
the six phases on the orbit are shown by dashed vertical lines.
\label{dissipation_fraction}}
\end{center}
\end{figure}
\newline\noindent{\bf The flow structure of bursting}\hspace{10pt} 
Figure~\ref{physical_space} shows the time-evolution of flow structures
along the homoclinic orbit at the six phases indicated in Fig.~\ref{phase_space}.
In the early stage of the evolution (phase a)
the spanwise standing-wave motion of the streak is enhanced,
so that the streamwise dependence, i.e. the three-dimensionality,
of the streak gradually becomes significant.
At the same time the streak grows in the wall-normal direction.
Such behaviour is a direct consequence of the linear instability of the UPO.
The eigenstructure for the instability is characterized by
disturbances of the streamwise velocity and vorticity
which are highly localized on the crest and the valley of the streak.\\
As time progresses,
the simultaneous spanwise oscillation and wall-normal growth
of the streak exceed a critical level for bursting.
There appear thin layers of extremely high vorticity
between the wall and the crest and the valley of the streak,
in which intense quasi-streamwise vortex tubes are generated
to induce high shear and thereby high energy dissipation (phase b).
The quasi-streamwise vortices of opposite signs of the streamwise
vorticity align following the oscillation of the streak in the
spanwise direction.
Then the streak is rapidly deformed, while the second pair of
quasi-streamwise vortices appears on the crest and valley,
and align with the first pair to form an array which
moves fast on the crest or valley in the spanwise
direction (phase c).
During this streak deformation and vortex generation
intense energy dissipation is observed in the
thin layers between the wall and the crest or the valley of the streak.
Streamwise vortices similar to those found in the regeneration cycle \cite{hamilton}
and the corresponding strong UPO \cite{kawa} also appear on the
flanks of the streak.
However, the streamwise vortices on the crest and valley of the
streak are not observed in the near-wall regeneration cycle,
and therefore we can say that these vortices are typical
flow structure of bursting with significant energy dissipation.
Actually, an inspection of the turbulent state of plane Couette flow
has shown that in bursting events of intense dissipation
there appear
similar vortex arrays on the crest and valley of the
highly grown streak, which are associated with strong
local dissipation.
As shown in Fig.~\ref{dissipation_fraction},
the intense local dissipation (represented by
green isosurfaces) observed around
the crest and valley of the 
streak
in Fig.~\ref{physical_space} (phases c and d)
contributes to the rapid increase of total energy
dissipation (see Fig.~\ref{phase_space}).
The identified high-dissipation regions occupy less than 10\% of the spatial
domain but account for over 40\% of the energy dissipation, implying that
strong energy dissipation in the bursting event can be attributed to
flow structures localized in the region around the crest and the valley of the
streak shown by the green surfaces in Fig. \ref{physical_space}.
As time goes on, the highly deformed streaks
are broken down (phase d), and then the streak and vortices decay
rapidly (phase e) as the flow finally returns to a quiescent state close to the gentle UPO
(phase f).\\
Transient approaches of a turbulence state to
relatively quiescent states of low energy input and dissipation are followed
by bursting events with intense energy dissipation,
as reported by Kawahara \& Kida~\cite{kawa}, who
suggested the interpretation of bursting
in terms of a two-way heteroclinic connection between the gentle UPO
and a UPO embedded in turbulent flow.
The present more rigorous analysis has
presented another interpretation, namely that
the whole process of the bursting, including
a transient approach to a quiescent state and a subsequent
highly dissipative event, is the manifestation of a
homoclinic orbit arising from the gentle UPO alone.\\
\noindent{\bf Conclusion}\hspace{10pt} The existence of an orbit homoclinic to this edge state  
implies that the geometry of the laminar-turbulent boundary is rather complex and we can expect to 
find a manageable approximation to it only locally. At the same time, it implies the existence of 
infinitely many UPOs which correspond to flows with arbitrarily many, arbitrarily long, near laminarization
events. It is natural then to think of turbulent shear flow as governed by a large chaotic attractor
which comprises both periodic orbits in the turbulent regime, which reproduce the regeneration
cycle \cite{kawa}, and periodic orbits which reproduce near-laminarization and bursting events.
Physically, these bursting events are very different from the regeneration cycle. The streamwise
vortices appear on the crest and valley of the streaks rather than on their flanks, and the larger
part of the energy dissipation takes place in the cycle around the vortices in the near-wall, high strain 
region.\\
The homoclinic solution presented here adds a new element to the elucidation of turbulence by means
of dynamical systems theory, namely that of temporal localization. Since it has been computed in a
minimal flow unit, in cannot capture spatial intermittency. Recently, several equilibrium and travelling
wave solutions exhibiting spatial localization have been found \cite{schneider10,*ehrenstein08}. A clear goal
for the near future is to find periodic or connecting orbits which combine the two forms of localization.

\bibliographystyle{natbib}
\bibliography{Couette_c2c}

\end{document}